\documentclass{Interspeech2024}

\usepackage{acronym}
\usepackage{cite}
\usepackage{subfig,balance}
\usepackage{tabularx}
\usepackage{amsmath,amssymb,amsfonts,mathtools}

\usepackage{pifont}%
\newcommand{\cmark}{\ding{51}}%
\newcommand{\xmark}{\ding{55}}%

\acrodef{TF}{Time-Frequency}
\acrodef{CLIP}{Contrastive Language-Image Pre-Training}
\acrodef{CLAP}{Contrastive Language-Audio Pre-training}
\acrodef{RIR}{Room Impulse Responses}
\acrodef{FFT}{Fast Fourier Transform}
\acrodef{FF}{Feed Forward}
\acrodef{AST}{Audio Spectrogram Transformer}
\acrodef{GMM}{Gaussian Mixture Model}
\acrodef{SVM}{Support Vector Machine}
\acrodef{DNN}{Deep Neural Networks}
\acrodef{CNN}{Convolutional Neural Networks}
\acrodef{CRNN}{Convolutional
Recurrent Neural Network}
\acrodef{ACE}{Acoustic
Characterization of Environments}

\interspeechcameraready 

\title{RevRIR: Joint Reverberant Speech and Room Impulse Response Embedding using Contrastive Learning with Application to Room Shape Classification}

\name[affiliation={1}]{Jacob}{Bitterman}
\name[affiliation={2}]{Daniel}{Levi}
\name[affiliation={2}]{Hilel Hagai}{Diamandi}
\name[affiliation={2}]{Sharon}{Gannot}
\name[affiliation={1}]{Tal}{Rosenwein}

\address{
  $^1$OrCam Technologies LTD, Israel\\
  $^2$Bar Ilan University, Israel}
  
\email{Jacobitterman@gmail.com, daniel.levi1@biu.ac.il, hagai.diamandi@yale.edu, sharon.gannot@biu.ac.il, tal.rosenwein@orcam.com}


\begin{document}

\maketitle

\begin{abstract}
    
This paper focuses on room fingerprinting, a task involving the analysis of an audio recording to determine the specific volume and shape of the room in which it was captured. While it is relatively straightforward to determine the basic room parameters from the \ac{RIR}, doing so from a speech signal is a cumbersome task.
To address this challenge, we introduce a dual-encoder architecture that facilitates the estimation of room parameters directly from speech utterances. During pre-training, one encoder receives the \ac{RIR} while the other processes the reverberant speech signal. A contrastive loss function is employed to embed the speech and the acoustic response jointly. In the fine-tuning stage, the specific classification task is trained. In the test phase, only the reverberant utterance is available, and its embedding is used for the task of room shape classification. The proposed scheme is extensively evaluated using simulated acoustic environments.

\end{abstract}

\section{Introduction}
Acoustic scene analysis has emerged as an active research field in recent years.\footnote{https://dcase.community/} In this study, we focus on the task of classifying speech utterances based on the volume and shape of the room in which they were recorded. Estimating room volume and shape has numerous applications, including virtual and augmented reality, acoustic simulation, and forensic audio analysis.

Several studies addressed this task. It is shown in \cite{moore2013roomprints,moore2014room} that the
\ac{RIR}, specifically the reverberation time per frequency band, encompasses relevant information for identifying a room, and the notion ``roomprint'' is introduced. 
Hence, it can be deduced that the analysis of the \ac{RIR} provides an indication of the room volume and shape. However, this analysis is not as straightforward if only the speech utterance is given. In \cite{shabtai2010room,shabtai2013towards}, a set of carefully designed features is extracted from both time- and frequency-domain representations of the \ac{RIR}. The features include the density of the early reflections, the decay rate of the \ac{RIR}, the number of modes, the kurtosis of the frequency response, and many more. When only speech signals are available, it is proposed to either apply a dereverberation step or detect abrupt speech stops. The actual room classification was done using a \ac{GMM} clustering of the input features. A set of reverberation-related features is proposed in \cite{murgai2017blind} for room classification from reverberant speech using abrupt speech stops. A trained regressor is used to infer the room volume. 
In \cite{baum2022environment}, a variant of the \ac{RIR} is first blindly estimated from the recorded speech, and then the reverberation time is inferred per frequency band. A \ac{GMM}-based classifier is then applied to infer the room volume. Several combinations of features and classifiers, including \ac{SVM} classifier, are compared in \cite{mascia2015forensic}.

In recent years, the power of \ac{DNN} was harnessed to improve the room volume classification results. In \cite{yu2020room}, room geometry is inferred from a single \ac{RIR} using \ac{CNN} architecture.
In \cite{genovese2019blind}, a \ac{CNN} architecture is used to map speech spectrogram to the room volume directly. 
In \cite{papayiannis2020end}, a \ac{CRNN} architecture with an attention mechanism is employed to address the room classification task from a speech signal directly. The algorithm is tested using real recordings from the  \ac{ACE} challenge \cite{eaton2016estimation}.

In this work, we propose leveraging the latent relationship between the \ac{RIR} and the corresponding reverberant speech through the application of a contrastive loss, allowing for the joint embedding of speech and acoustic responses. This architecture aims to capture the acoustic response's fundamental characteristics, facilitating various downstream tasks. We refer to our proposed scheme as `RevRIR'. Its effectiveness is extensively evaluated using simulated data.


\section{Method}

\begin{figure*}[t]
  \centering
  \includegraphics[width=\linewidth]{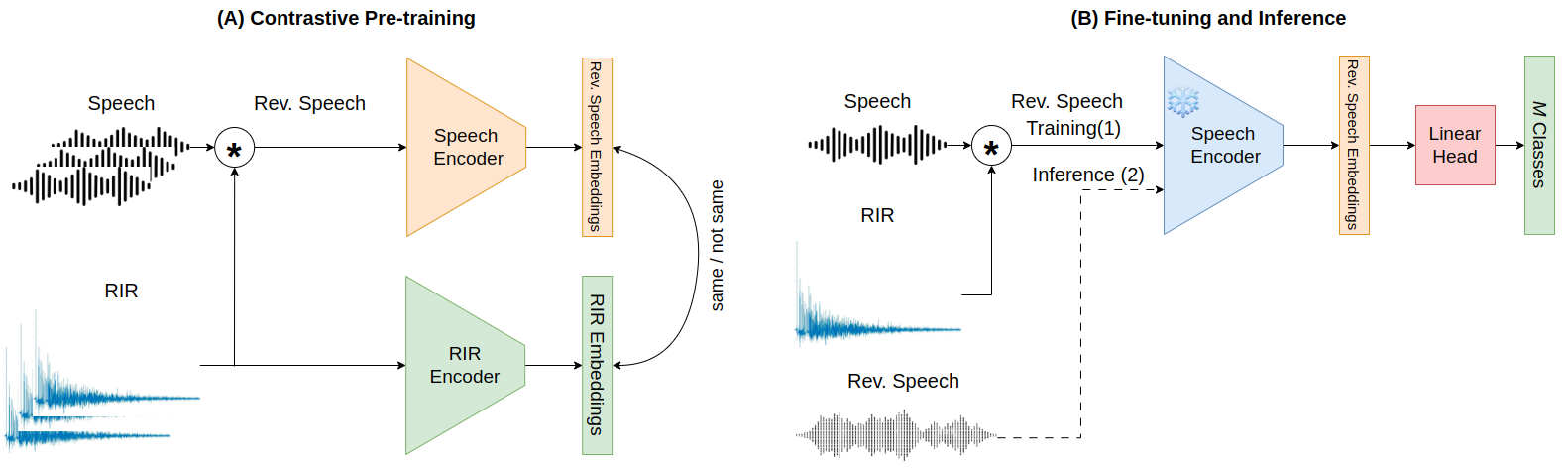}
   \addtolength{\belowcaptionskip}{-16pt}
\addtolength{\abovecaptionskip}{-10pt}
  \caption{System overview. (A) Pre-training: Using contrastive learning, two separate encoders are trained on the reverberated audio and the \ac{RIR}. (B) Fine-tuning: (1) freezing the reverberated speech encoder while training a classification head for the downstream task (e.g., classification to one of 110 rooms). In inference (2) a reverberated speech can be classified using the network.}
  \label{fig:CARIR_model}
\end{figure*}

Our main goal is to learn to relate room acoustics and reverberant utterances (under the same acoustic conditions) using two separate encoders and a contrastive learning objective to bring them into a joint embedding space. This joint space can later serve as a seed for downstream tasks, such as room fingerprinting. Thus, our method requires pre-training and fine-tuning steps, as shown in Fig.~\ref{fig:CARIR_model}. Our method is inspired by the \ac{CLIP} \cite{CLIP} and \ac{CLAP}
 \cite{CLAP} models.

\subsection{Pre-Training}
The pre-training step learns the joint embedding space as shown in Fig.~\ref{fig:CARIR_model}(A). 
Let $h_l(t)$ be an \ac{RIR} from class \(l \in M\) where $M$ is the number of room classes. Let $s(t)$ be an utterance. We generate the reverberated utterance $x(t)$ by convolving the utterance with the \ac{RIR}:
\begin{equation}
x(t)=\{s\ast h_l\}(t) \in \mathbb{R}^T,
\label{equation:eq_reverberation}
\end{equation}
where $\ast$ denotes the convolution operator, and $T$ is the sequnce length.
The reverberated utterance $x(t)$ is encoded using an \ac{AST} model \cite{gong21b_interspeech} $f_{\textrm{AST}}(\cdot)$ to produce the speech embeddings:
\begin{equation}
    \mathbf{e}_1=f_{\textrm{AST}}(x(t)) \in \mathbb{R}^d,
\end{equation}
where $d$ is the embedding dimension. The \ac{RIR} is encoded using \(f_{\textrm{RIR}}(\cdot)\), a sequence of feed-forward layers with ReLU activations. \(f_{\textrm{RIR}}(\cdot)\) is applied to the log magnitude of the \ac{FFT} of the \ac{RIR} to produce the \ac{RIR} embeddings $\mathbf{e}_2$:
\begin{equation}
\label{equation:FFT}
\mathbf{e}_2=f_{\textrm{RIR}}(20 \log_{10}(|\mathfrak{F}(h(t)|))\in \mathbb{R}^d,
\end{equation}
where \(\mathfrak{F}(\cdot)\) is the \ac{FFT} operator.
The embeddings are then normalized to have a unity norm. 
Define the  matrices  $\mathbf{E}_1,\mathbf{E}_2 \in \mathbb{R}^{B\times d}$, which comprise a minibatch of size $B$ of stacked embedding vectors ${\mathbf{e}_i^j}, i \in \{1, 2\}
, j \in\{1,..., B\}$ at each row.
The similarity between the normalized embedding vectors of the speech utterances and the \acp{RIR} is calculated using the dot product, followed by a softmax operator: 
\begin{equation}
    \textrm{SMDP}[\mathbf{e}_1^i, \mathbf{e}_2^j,] = \frac{\exp((\mathbf{e}_1^i)^\top  \mathbf{e}_2^j/\tau)}{\sum\limits_{k=1}^{B}\exp((\mathbf{e}_1^i)^\top  \mathbf{e}_2^k/\tau)} \in \mathbb{R}, 
\label{eq:soft_max_dot_product}
\end{equation}

where $\tau \in (0, 1]$ is a trainable temperature hyperparameter.
We then aggregate the similarity score for each entry to obtain: 
\begin{equation}
    \textrm{loss}[\mathbf{e}_1^i, \mathbf{E}_2] = -\sum\limits_{j\in \mathcal{N}_i}\frac{1}{|\mathcal{N}_i|}\log(\textrm{SMDP}[\mathbf{e}_1^i, \mathbf{e}_2^j,] ) ,
  \label{eq:loss_per_room_soft_max_dot_product}
\end{equation}

where $\mathcal{N}
_i$ is the subset of indices within the batch that belong to the same class $i$. Eq.~\eqref{eq:loss_per_room_soft_max_dot_product} encourages a similarity value of 1 for utterances that were reverberated using the \ac{RIR} from the same class within the batch and a similarity value of 0 otherwise. We then average the loss over all instances in the batch:
\begin{equation}
    \textrm{loss}[\mathbf{E}_1, \mathbf{E}_2] = \frac{1}{B}\sum\limits_{i=1}^{B} \textrm{loss}[\mathbf{e}_1^i, \mathbf{E}_2].  \label{eq:loss_final_soft_max_dot_product}
\end{equation}

Finally, we average two loss terms, one that compares each \ac{RIR} with all reverberated utterances and the second that compares each reverberated utterance to all \acp{RIR}:
\begin{equation}
    \mathcal{L} = \frac{\mathrm{loss}[\mathbf{E}_1, \mathbf{E}_2] + \mathrm{loss}[\mathbf{E}_2, \mathbf{E}_1]}{2}.
\label{eq:total_loss_final_soft_max_dot_product}
\end{equation}
As shown in \eqref{eq:total_loss_final_soft_max_dot_product}, the contrastive loss was adjusted such that \acp{RIR} and reverberated speech from the same class will have similar embeddings, even if the utterance was not convolved with the same exact \ac{RIR}, and vice versa. This encourages the embeddings to capture the acoustics of the utterance, ignoring the speaker attributes, spoken content, etc.

\subsection{Fine-Tuning}
After the model has been pre-trained, we add a trainable \ac{FF} linear head on top of either one of the encoders $f_{\textrm{AST}}(\cdot)$ or $f_{\mathrm{RIR}}(\cdot)$. In this work, we focus on audio fingerprinting to classify the reverberated utterance into one of the $M$ classes using the categorical cross-entropy loss. 

There are two options for fine-tuning. In the first, we train only the \ac{FF} head to solve the downstream task while freezing the encoder, as depicted in Fig.~\ref{fig:CARIR_model}(B1). In this scenario, since the encoder is frozen, either one of the encoders can be utilized during inference. Subsequently, the embeddings are projected using the \ac{FF} head. An illustration of the inference process is provided in Fig.~\ref{fig:CARIR_model}(B2), where the frozen $f_{\textrm{AST}}(\cdot)$ encoder is applied to an utterance, its embedding is computed, and the resulting embedding is passed to the classification head to obtain the utterance's room dimensions and volume.
Alternatively, both the encoder and the \ac{FF} head can be jointly trained. In this case, we are restricted to only applying the selected encoder during inference.

\section{Algorithm and Experimental Setup}
In this section, we will describe the datasets, the models' configurations, and training procedures.
\subsection{Speech Datasets} \label{sec:Speech_dataset}

Clean and non-reverberant utterances were drawn from a subset of 36K hours from LibriVox\footnote{http://www.openslr.org/94/} as described in \cite{AWP}, and from LibriSpeech-train \cite{LibriSpeech} datasets. LibriSpeech dev-clean was chosen as the validation set, as it contains unseen speakers and spoken content w.r.t.~the training set.
The audio was resampled to 8KHz, 16 bits/sample, and $T$ was set to $\leq 10$~seconds.

\subsection{\acfp{RIR}}  \label{sec:RIR_dataset}
The training and validation \acp{RIR} were generated using the \ac{RIR} generator Python package.\footnote{https://pypi.org/project/rir-generator/} We chose $M=110$ rooms derived from three room types as described in Table~\ref{tab:rooms}.
\begin{table}[th]
 \addtolength{\belowcaptionskip}{-6pt}
\addtolength{\abovecaptionskip}{-4pt}
  \caption{Room dimensions' range is in the form of [min, max, hop]. There are 16 small rooms, 52 large rooms, and 42 halls, resulting in $M=110$ room classes.}
  \label{tab:rooms}
  \centering
  \resizebox{0.95\columnwidth}{!}{
    \begin{tabular}{llll}
    \toprule
    \multicolumn{1}{c}{\textbf{}} & 
                                         \multicolumn{1}{c}{\textbf{Small}} & 
                                         \multicolumn{1}{c}{\textbf{Large}} & 
                                         \multicolumn{1}{c}{\textbf{Hall}} \\
    \midrule
    Width [m]                      & [1.5, 3.5, 1.0] & [6.0, 13.0, 1.0] & [1.0, 3.0, 1.0]             \\
    Depth [m]                      & [2.5, 4.5, 1.0]  & [6.0, 12.0, 2.0] & [7.0, 13.0, 1.0]              \\
    Height [m]                      & [2.5, 3.0, 0.5] & [2.5, 3.5, 1.0] & [2.5, 3.5, 1.0]       \\
    \bottomrule
    
  \end{tabular}
  }
  
\end{table}
A total of 550K \acp{RIR} were generated, split uniformly between the $M=110$ room classes.
The parameters controlling the \acp{RIR} are: 1) speed of sound  $c=343$~[m/s], and 2) reflection coefficient of all room facets $\beta \in [0.88, 0.9]$, 3) sampling rate 8KHz, 4) minimal distance from the source to the walls 0.5~m, and 5) minimal distance between the source and the microphone 0.5~m. All \acp{RIR} were zero-padded to 4096. 
For the validation set, we chose a subset of these \acp{RIR}, ensuring that they are not included in the training data, although they may originate from familiar rooms.

\subsection{Configurations, Training, and Hyperparameters}
\textbf{Pre-training:} For the Reverberant Speech encoder we took a pretained  \ac{AST} mode\footnote{https://github.com/lucidrains/musiclm-pytorch} \cite{gong21b_interspeech} with 85M parameters, which outputs $\mathbf{e}_1 \in \mathbb{R}^{768}$, i.e., the embedding dimension $d$ was set to 768.
The parameters of the \ac{RIR} encoder were initialized using Xavier initialization.\footnote{Katanforoosh \& Kunin, ``Initializing neural networks,'' deeplearning.ai, 2019.} 
For the \ac{RIR} encoder, we used a sequence of four feed-forward blocks (with corresponding output dimensions of 3264, 2432, 1600, and 768), each consisting of Linear Layer, Relu activation, and BatchNorm, with a total of 26M parameters.
The output of the \ac{RIR} encoder has the same dimensions, i.e., $\mathbf{e}_2 \in \mathbb{R}^{768}$.
We trained the encoders for four epochs for two days on a single Nvidia RTX 2080-Ti GPU using a linear scheduler with a warmup ratio of 0.05, a learning rate of 1e-5, and the AdamW optimizer. The batch size was set to $B=55$. 

\noindent\textbf{Fine-tuning:} 
During the fine-tuning stage, we utilized the encoder from the pre-training phase, followed by a straightforward Linear head with an input dimension of $d=768$ and an output dimension equal to the number of classes, $M=110$ in our case.  

In this study, our primary focus is inferring the room shape from reverberated speech. However, we will also assess the classification performance using either the reverberated speech or directly the \ac{RIR}. For both cases, we will utilize the respective encoders from the pre-training stage, \emph{with} or \emph{without} freezing their weights. For the fine-tuning stage, we trained the model for 50 epochs on a single Nvidia RTX 2080-Ti GPU for two days, using a polynomial learning rate scheduler with a power of 0.1, a learning rate of 1e-4, and the AdamW optimizer. The batch size, in this case, was set to $B=100$. These hyperparameters were chosen based on the loss over the training set.
All four models were trained using the same hyper-parameters and pre-trained model.

\subsection{Evaluation Metrics}
To test the effectiveness of our framework, we evaluated the fine-tuned models using a standard implementation of Top-1 Accuracy on two tasks: 1) all 110 room classes or 2) only 3 room types (see Table~\ref{tab:rooms}). 
The latter evaluation was implemented by first applying the 110-room classifier and then aggregating the rooms into three room types.
The evaluation data differed from the training samples as described in sections \ref{sec:Speech_dataset}, \ref{sec:RIR_dataset}; specifically, the same \ac{RIR} could not appear in both the training and test datasets, although they might originate from the same rooms.  

We applied several evaluation procedures: 1) the quality of the pre-training stage was visualized using t-SNE \cite{van2008visualizing}; 2) top-1 accuracy was used to assess the various training variants of the 110-room classification task and to compare them with a baseline method; and 3) confusion matrices were used for the 3-room type classification task.    

\subsection{Competing Method}

To further evaluate the performance of our model, we compared its performance against a standard classification framework. For each \ac{RIR}, we calculated a 30-feature vector and used it as input to the classifier. The \ac{RIR}-based parameters are listed in \cite{shabtai2013towards} (Table 1). The frequency domain parameters were calculated from the room transfer function, namely the Fourier transform of the \ac{RIR}, split into five-octave bands,  $[k \times 50 ,k \times 200]$~Hz,\, $1\leq k \leq 5$.

We substituted the original classifier in \cite{shabtai2013towards} with a classification network comprising four fully connected layers with output dimensions of 65, 90, 100, and 110, following ReLU activation functions, except the last layer. The network was trained using a linear scheduler with a warmup ratio of 0.1, a learning rate of 1e-3, and a cross-entropy loss function, implementing a dropout rate of 0.2 during the training stage over 100 epochs.

\section{Results} \label{sec:Results}
\begin{figure}[t]
  \centering
\includegraphics[width=0.9\linewidth]{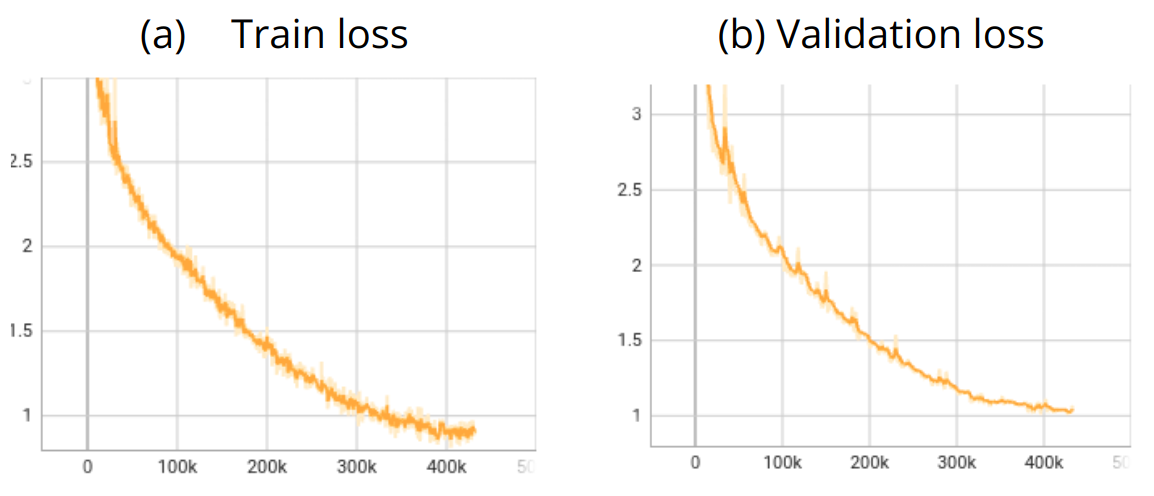}
  \addtolength{\belowcaptionskip}{-18pt}
\addtolength{\abovecaptionskip}{-5pt}
  \caption{(a) Training and (b) Validation losses during the pre-training stage.}
  \label{fig:pretraining_loss}
\end{figure}

In this section, the performance of the RevRIR scheme is evaluated and compared with the competing method. Figure~\ref{fig:pretraining_loss} depicts the loss values of the training and validation sets during the pre-training stage. Evidently, the pre-training is stable, converges to a minimal value, and no overfitting is observed.
\subsection{110-Room Classification} \label{sec:finetuning_results}
\noindent\textbf{Pre-training:} 
\begin{figure*}[t]
  \centering
\includegraphics[width=0.9\linewidth]{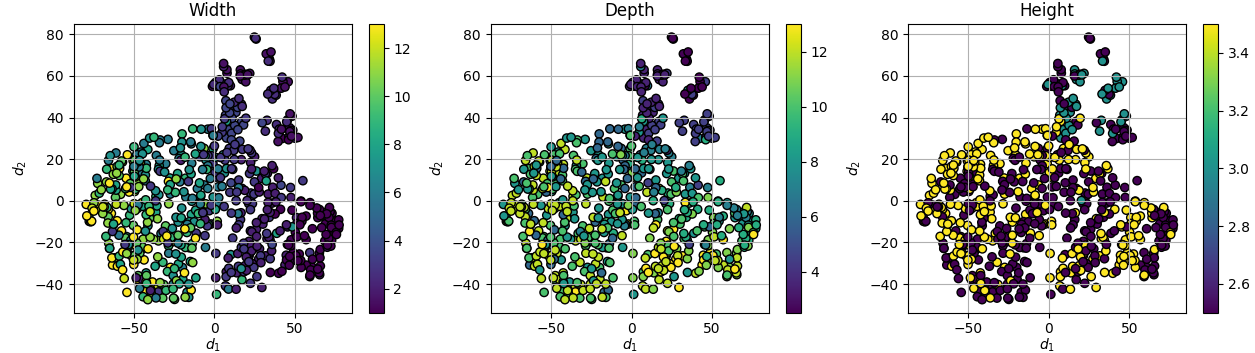}
 \addtolength{\belowcaptionskip}{-10pt}
\addtolength{\abovecaptionskip}{-4pt}
  \caption{Visualizing the embedding space of validation samples projected using t-SNE. The left, middle, and right graphs show the distribution of projected embeddings, colored by their ground truth width, depth, and height values, respectively. It is clearly observed that the features were projected according to the room dimension and that it is content- and speaker-independent.}
  \label{fig:embedding_space}
\end{figure*}
Figure~\ref{fig:embedding_space} depicts the t-SNE visualization of reverberated speech samples from the validation set, with 50 different speakers and five different \acp{RIR} from each room type. Overall, 12,500 utterances are visualized. The projected embeddings are colored by their ground truth width, depth, and height values.  It is clear that the embedding space is speaker-independent and also content-independent. Moreover, it conveys semantic meaning regarding width and depth, where progressing from left to right signifies an increase in room width, and from top to bottom indicates an increase in room depth. However, this semantic order is not noticeable concerning height, possibly due to the relatively limited range of heights compared to the other dimensions.

\noindent\textbf{Fine-tune Stage:} Table~\ref{tab:fine_tuning} depicts the Top-1 Accuracy results of the fine-tuned models compared to random guess (1/110=0.9\%) and the baseline model utilizing the \ac{RIR} features (adapted from \cite{shabtai2013towards}). 
We can see a clear potential for the fine-tuned models as they are significantly above the random guess. 
For statistical significance, we trained the models using four different seeds. We found a standard deviation of 0.21\% and 0.12\% on the Reverberated Speech  and the \ac{RIR} validation set, respectively.
Evidently, the models trained with the \ac{RIR} encoder achieve higher accuracy on the \acp{RIR} validation set. 

The model fine-tuned with the frozen speech encoder outperforms the one fine-tuned on the frozen RIR encoder by 3\% in accuracy (40\% vs. 37\%) on the Reverberated Speech set, despite its lower accuracy on the RIRs set (66\% vs. 83\%). However, the models fine-tuned without freezing the encoder achieve significantly higher accuracy on the relevant validation set.

It is important to note that although the model's performance with a frozen encoder may be inferior, in many cases, such as when there are several downstream tasks, freezing the encoder has a clear advantage, as it requires calculating the embeddings only once instead of once per downstream task. 
\begin{table}[htp]
 \addtolength{\belowcaptionskip}{-6pt}
\addtolength{\abovecaptionskip}{-4pt}
  \caption{Top-1 Accuracy of the fine-tuned models on unseen speakers (LibriSpeech Dev clean) dataset, and unseen \ac{RIR} (from the 110 known rooms), compared to random guess and baseline model.}
  \label{tab:fine_tuning}
  \centering
  \resizebox{0.95\columnwidth}{!}{
    \begin{tabular}{lccc}
    \toprule
     \multicolumn{1}{c}{\shortstack[t]{\textbf{Training} \\ \textbf{Method}}} & 
     \multicolumn{1}{c}{\shortstack[t]{\textbf{Freezed} \\ \textbf{Encoder?}}} & 
     \multicolumn{1}{c}{\shortstack[t]{\textbf{\acp{RIR} Set} \\ Top-1 [\%]}} & 
     \multicolumn{1}{c}{\shortstack[t]{\textbf{Rev. Speech Set} \\ Top-1 [\%]}} \\
    
    \midrule
    Random         & --- & $0.9$   & $0.9$          \\
    \midrule
    \ac{RIR} feat. based on  \cite{shabtai2013towards} & ---   & $31.3$    & ---               \\
    \midrule
    RIR            & \cmark   & $83$    & $37$              \\
    RIR            & \xmark   & $99.9$  & ---               \\
    Rev. Speech         & \cmark   & $66$    & $40$           \\
    Rev. Speech         & \xmark   & ---     & 60             \\
    \bottomrule
  \end{tabular}
  }
\end{table}
\subsection{Classification of 3-Room Types}
We now turn to the task of room type classification. 
As explained in Table \ref{tab:rooms}, the rooms are categorized into three types based on their volume and shape: small, large, and hall. 
To obtain these results, we first classify the signals (or \acp{RIR}) into 110 rooms and then merge the classes into three room-type categories. 
 We evaluate and compare three models by analyzing the confusion matrices: 1) Table~\ref{tab:Audio fine tune}: fine-tuned with a frozen speech encoder , 2) Table~\ref{tab:RIR fine tune}: fine-tuned with a frozen \ac{RIR} encoder, and 3) Table~\ref{tab:RIR baseline model}: the baseline method utilizing features extracted from the \ac{RIR}, as explained in \cite{shabtai2013towards} . 

 We observe that both variants of RevRIR perform exceptionally well on this task, achieving accuracies of 95.4\% and 99.6\%, respectively. Moreover, it is evident that RevRIR significantly surpasses the baseline method, achieving an accuracy of 85.8\%.
\begin{table}[h!]
 \addtolength{\belowcaptionskip}{-6pt}
\addtolength{\abovecaptionskip}{-4pt}
  \caption{Confusion Matrix of the model fine-tuned with frozen Speech encoder on unseen speakers (LibriSpeech Dev clean) dataset, and unseen \acp{RIR} (albeit seen rooms), collapsed to 3 room types.}
  \label{tab:Audio fine tune}
  \centering
  \resizebox{0.75\columnwidth}{!}{
    \begin{tabular}{cccc}
    \toprule
    \multicolumn{1}{c}{\textbf{GT / Prediction}} & 
                                         \multicolumn{1}{c}{\textbf{Small}} & 
                                         \multicolumn{1}{c}{\textbf{Large}} & 
                                         \multicolumn{1}{c}{\textbf{Hall}} \\
    \midrule
    \textbf{Small}                  & 0.938 & 0.    & 0.062       \\
    \textbf{Large}                  & 0.    & 0.968 & 0.032       \\
    \textbf{Hall}                   & 0.02  & 0.022 & 0.958       \\
    \bottomrule
  \end{tabular}
  }
\end{table}

\begin{table}[h!]
 \addtolength{\belowcaptionskip}{-6pt}
\addtolength{\abovecaptionskip}{-4pt}
  \caption{Confusion Matrix of the model fine-tuned with frozen \ac{RIR} encoder on unseen speakers (LibriSpeech Dev clean) dataset, and unseen \acp{RIR} (albeit seen rooms), collapsed to 3 room types.}
  \label{tab:RIR fine tune}
  \centering
  \resizebox{0.75\columnwidth}{!}{
    \begin{tabular}{cccc}
    \toprule
    \multicolumn{1}{c}{\textbf{GT / Prediction}} & 
                                         \multicolumn{1}{c}{\textbf{Small}} & 
                                         \multicolumn{1}{c}{\textbf{Large}} & 
                                         \multicolumn{1}{c}{\textbf{Hall}} \\
    \midrule
    \textbf{Small}                  & 0.997 & 0.    & 0.003       \\
    \textbf{Large}                  & 0.    & 0.997 & 0.003       \\
    \textbf{Hall}                   & 0.001 & 0.004 & 0.995       \\
    \bottomrule
  \end{tabular}
  }
\end{table}

\begin{table}[h!]
 \addtolength{\belowcaptionskip}{-6pt}
\addtolength{\abovecaptionskip}{-8pt}
  \caption{Confusion Matrix of the RIR features model (adapted from \cite{shabtai2013towards}) on unseen speakers (LibriSpeech Dev clean) dataset, and unseen \acp{RIR} (albeit seen rooms), collapsed to 3 room types.}
  \label{tab:RIR baseline model}
  \centering
  \resizebox{0.75\columnwidth}{!}{
    \begin{tabular}{cccc}
    \toprule
    \multicolumn{1}{c}{\textbf{GT / Prediction}} & 
                                         \multicolumn{1}{c}{\textbf{Small}} & 
                                         \multicolumn{1}{c}{\textbf{Large}} & 
                                         \multicolumn{1}{c}{\textbf{Hall}} \\
    \midrule
    \textbf{Small}                  & 0.922 & 0.    & 0.023       \\
    \textbf{Large}                  & 0.    & 0.823 & 0.147       \\
    \textbf{Hall}                   & 0.078 & 0.177 & 0.83       \\
    \bottomrule
  \end{tabular}
  }
\end{table}

\section{Summary and Discussion}
We have presented a model to infer room volume and shape from single-microphone speech recordings. A contrastive loss function is employed in the proposed architecture to embed the speech and the acoustic response jointly. After the model has been pre-trained, we add a trainable \ac{FF} linear head on top of either one of the encoders to perform the actual classification task. The model was tested on simulated data, exhibiting excellent classification capabilities. This work serves as a proof of concept for the ability to conduct such tasks. It can be readily applied to other tasks, such as determining if two recordings originated from the same room, which has clear applications in the audio forensics domain. In the future, we plan to evaluate the performance with real room recordings recorded in rooms with various shapes and structures.  

\balance
\bibliographystyle{IEEEtran}
\bibliography{mybib}

\end{document}